\documentclass[%
 reprint,
 amsmath,amssymb,
 aps,
]{revtex4-2}

\usepackage{graphicx}
\usepackage{dcolumn}
\usepackage{bm}
\usepackage{natbib}
\usepackage{booktabs}

\begin{document}

\preprint{APS/123-QED}

\title{Phase Plane Analysis of Firing Patterns in the Adaptive Exponential Integrate-and-Fire Model
}

\author{Wu-Fei Zhang}
 \altaffiliation{The School of Physical Science and Technology, Lanzhou University,Lanzhou 730000, China}
 \email{zhangwf2023@lzu.edu.cn}

\begin{abstract}
The Adaptive Exponential Integrate-and-Fire (AdEx) model is a simplified framework that effectively characterizes neuronal electrical activity. The aim of this paper is to employ phase plane analysis to systematically investigate diverse firing patterns generated by the AdEx model under varying parametric conditions. We first introduce the fundamental equations and parameter configurations of the AdEx model to numerically simulate the six representative firing patterns in the AdEx model. And then we use phase plane analysis to explore the dynamic mechanism of these firing patterns under different input currents and parametric conditions. Our findings demonstrate that the AdEx model can simulate multiple firing patterns, including Tonic Spiking, Adapting, Initial Bursting, Busting, Transient Spiking and Delayed Spiking firing patterns. These results not only advance the understanding of complex electrophysiological phenomena in neurons but also provide theoretical foundations for applications in many fields like neuromorphic computing and brain-computer interfaces.
\end{abstract}

\maketitle

\tableofcontents

\section{Introduction}

Neurons, as the fundamental functional units of the nervous system, exhibit membrane potential generation mechanisms that constitute one of the core research questions in neuroscience. In the 1950s, Hodgkin and Huxley conducted a series of pioneering experiments that led to the Hodgkin-Huxley (H-H) model\cite{hodgkin1952components}\cite{hodgkin1952currents}\cite{hodgkin1952dual}\cite{hodgkin1952measurement}\cite{hodgkin1952quantitative}. This model represented the first mathematical framework capable of precisely describing neuronal action potentials, successfully elucidating the physiological mechanisms underlying action potential generation and establishing a solid theoretical foundation for neuroscience. Subsequently, the H-H model has been extensively applied in neurophysiology, brain science, and related fields, becoming a classical framework for studying neuronal electrophysiological properties.However, practical applications of the H-H model face significant challenges. To accurately simulate the intricate details of action potential generation, the model requires small numerical integration steps while incorporating multiple complex dynamic variables. This substantially increases computational costs, limiting its utility in large-scale neural network simulations. Moreover, from the perspective of overall neural system functionality, researchers often prioritize interactions between neurons over detailed mechanisms of individual action potentials.\cite{brainpy} Consequently, simplifying neuronal models becomes necessary in many contexts. Addressing these needs, researchers have developed simplified neuron models that balance biological plausibility and computational efficiency. \cite{jolivet2004generalized}\cite{brainpy}By strategically omitting certain details, these models reduce computational complexity while effectively capturing critical electrophysiological characteristics of neurons, thereby providing more efficient and practical tools for neuroscience research.  

The Adaptive Exponential Integrate-and-Fire (AdEx) model, proposed by Brette and Gerstner in 2005\cite{brette2005adaptive}, represents a significant advancement in computational neuroscience. By incorporating a nonlinear term, the model more accurately reflects real neuronal membrane potential dynamics. It further introduces an adaptation current \(\omega\) to characterize neuronal adaptive properties. These features enable the AdEx model to capture complex dynamic behaviors under varying conditions while maintaining computational efficiency. In recent years, the AdEx model has gained widespread application across neuromorphic computing, brain-computer interfaces, and neurological disorder research due to its unique advantages.\cite{indiveri2011neuromorphic}\cite{merolla2014million} Nevertheless, as research progresses, deeper exploration of the model's dynamic characteristics across parameter spaces becomes increasingly crucial. Phase plane analysis serves as a powerful tool for visualizing long-term behaviors and stability in dynamical systems, providing robust support for uncovering the intrinsic mechanisms of the AdEx model. Therefore, this study employs phase plane analysis to systematically investigate diverse firing patterns and their underlying dynamic mechanisms in the AdEx model under varying parametric conditions, aiming to establish a deeper theoretical foundation for related research fields.  

The primary aims of this paper are as follows:\begin{itemize}
    \item  Systematic introduction to the fundamental equations and parametric configurations of the AdEx model
    \item  Comprehensive exploration of diverse firing patterns, including regular spiking, burst firing, adaptive firing, chaotic discharges, and oscillatory activity  
    \item  Revelation of dynamic mechanisms governing firing patterns under varying input currents and parametric conditions through phase plane analysis  
\end{itemize}  
  And the structure of this paper is as follows. Section II introduces the fundamental equations and parametric configurations of the AdEx model. Section III investigates the model's dynamic behaviors under varying conditions through phase plane analysis. Section IV concludes the study and discusses potential applications of the findings.

\section{The Adaptive Exponential Integrate-and-Fire (AdEx) Model and its Firing patterns}

\subsection{The Adaptive Exponential Integrate-and-Fire (AdEx) Model}

The core of the AdEx model consists of two differential equations governing neuronal dynamics\cite{brainpy}:
\begin{align}
    \tau \frac{dV}{dt} &= - (V - V_{rest}) +  \Delta_T e^{\frac{V - V_T}{\Delta_T}}  - Rw + I \label{adex1}\\
    \tau_w \frac{dw}{dt} &= a(V - E_L) + b\tau_w \underset{t(f)}{\Sigma} \delta(t - t(f)) \label{adex2}
\end{align}
In the equations, \( w \) is referred to as the adaptation current. Equation 
\ref{adex1} describes the dynamics of the membrane potential V in the AdEx model, and shares similarities with the membrane potential in the Exponential Integrate-and-Fire (ExpIF) model \cite{fourcaud2003spike}\cite{brainpy}:
\begin{equation}
    \tau \frac{dV}{dt} = - (V - V_{rest}) + \Delta_T e^{\frac{V-V_T}{\Delta_T}} + RI
\end{equation},
but includes an additional term \(-Rw\) on its right-hand side, explicitly indicating that the membrane potential is modulated by the adaptation current \( w \).Equation \ref{adex2} governs the dynamics of \( w \) through three interconnected mechanisms: The first component \( a(V - V_{\text{rest}}) \) represents voltage-dependent positive regulation, where elevated membrane potentials accelerate the development of neuronal adaptation. The second term \(-w\) constitutes an exponential decay process. The third component describes spike-triggered adaptation through \( \frac{b}{\tau_w} \sum_{f} \delta(t - t{(f)}) \), where the Dirac delta function \( \delta(t - t{(f)}) \) integrates to unity precisely at each spike time \( t{(f)} \). The \( \tau_w \) factor cancels with the left-hand side coefficient, resulting in an instantaneous increment of \( w \) by magnitude \( b \) following every action potential. This integrated mechanism ensures each spike produces a fixed enhancement in neuronal adaptation. 

Finally, Equation \begin{equation}
    if \quad V > \theta, V_{reset} \to V \quad last \quad t_{ref}
\end{equation}
explicitly defines the membrane potential reset conditions following action potential generation, completing the adaptive feedback loop essential to the AdEx model's dynamics.

\subsection{Firing Patterns in the The Adaptive Exponential Integrate-and-Fire (AdEx) Model}

Neurons across distinct brain regions and of varying types exhibit diverse firing patterns. Based on the temporal intervals between action potentials under constant stimulation, neuronal discharge patterns can be categorized into four primary modes:  
\begin{itemize}
    \item \textbf{Tonic Spiking}:Consistent inter-spike intervals throughout stimulation  
    \item \textbf{Adapting}:Progressively increasing inter-spike intervals that stabilize asymptotically  
    \item \textbf{Bursting}:Periodic high-frequency spike clusters separated by extended silent periods  
    \item \textbf{Irregular spiking}: Stochastic discharge patterns without discernible temporal organization  
\end{itemize}

This classification system characterizes steady-state firing behavior following stimulus onset. However, many neurons display distinct initial response patterns during stimulus transition due to significant divergence between baseline and activated states, requiring finite transition time to reach equilibrium. Initial response dynamics manifest in three fundamental patterns:  
\begin{itemize}
    \item \textbf{Classic Spiking}: Initial response mirrors steady-state discharge characteristics.
    \item \textbf{Initial Bursting}: Transient high-frequency spike clusters exceeding steady-state firing rates.
    \item \textbf{Delayed Spiking}: Protracted subthreshold oscillations preceding first action potential.
\end{itemize}

Theoretically, the combinatorial space of initial (3 patterns) and steady-state (4 patterns) classifications yields 12 distinct firing pattern combinations, all empirically observed in electrophysiological recordings.\cite{markram2004_Interneurons_of_the_neocortical_inhibitory_system}  

\begin{table}[h]
    \centering
    \caption{Invariant parameters for the six representative firing patterns}
    \label{table2}
    \begin{tabular}{ccccc}
    \toprule
        \textbf{$V_{rest}$} & \textbf{$\theta$} & \textbf{$V_{T}$} & \textbf{$\Delta_T$} & R  \\
    \midrule
         -70mV & 0mV & -50mV & 2 & $0.5m\Omega$ \\ 
    \bottomrule
    \end{tabular}
\end{table}

\begin{table*}[ht]
\centering
\caption{The input currents and parametric conditions of multiple firing patterns in the AdEx model\cite{brainpy}}
\label{table1}
\begin{tabular}{ccccccc}
\toprule
\textbf{Firing Patterns} & \textbf{$\tau$(ms)} & \textbf{$\tau_w$(ms)} &\textbf{a}  & \textbf{b} & \textbf{$V_{reset}$(mV)} & \textbf{I(mA)}  \\
\midrule
\textbf{Spiking pattern} & 20 & 30 & 0 & 60 & -55 & 65 \\
\textbf{Adapting pattern} & 20 & 100 & 0 & 5 & -55 & 65 \\
\textbf{Initial Busting pattern} & 5 & 100 & 0.5 & 7 & -51 & 65  \\
\textbf{Busting pattern} & 5 & 100 & -0.5 & 7 & -51 & 65  \\
\textbf{Transient Spiking pattern} & 10 & 100 & 1 & 10 & -60 & 55  \\
\textbf{Delayed Spiking Spiking pattern} & 5 & 100 & -1 & 5 & -60 & 25  \\
\bottomrule
\end{tabular}
\end{table*}

\begin{figure}
    \centering
    \includegraphics[width=1.0\linewidth]{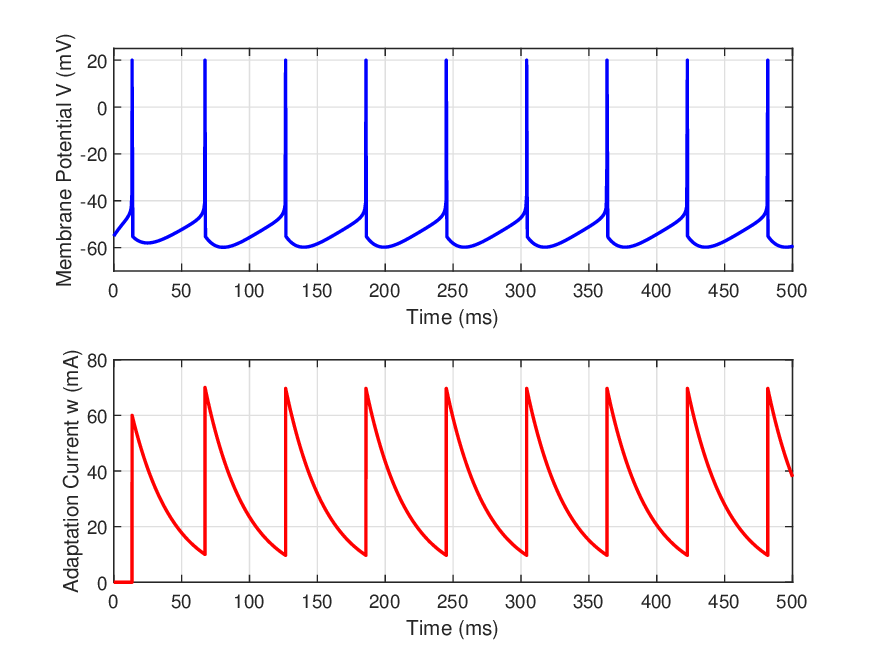}
    \caption{Simulation results of the Tonic Spiking pattern in the AdEx model. The upper figure shows the variation of the membrane potential \( V \) over time in the Tonic Spiking pattern, while the lower figure shows the variation of the adaptation current \( \omega \) over time in the  Tonic Spiking pattern.}
    \label{fig:Spiking pattern}
\end{figure}
\begin{figure}
    \centering
    \includegraphics[width=1.0\linewidth]{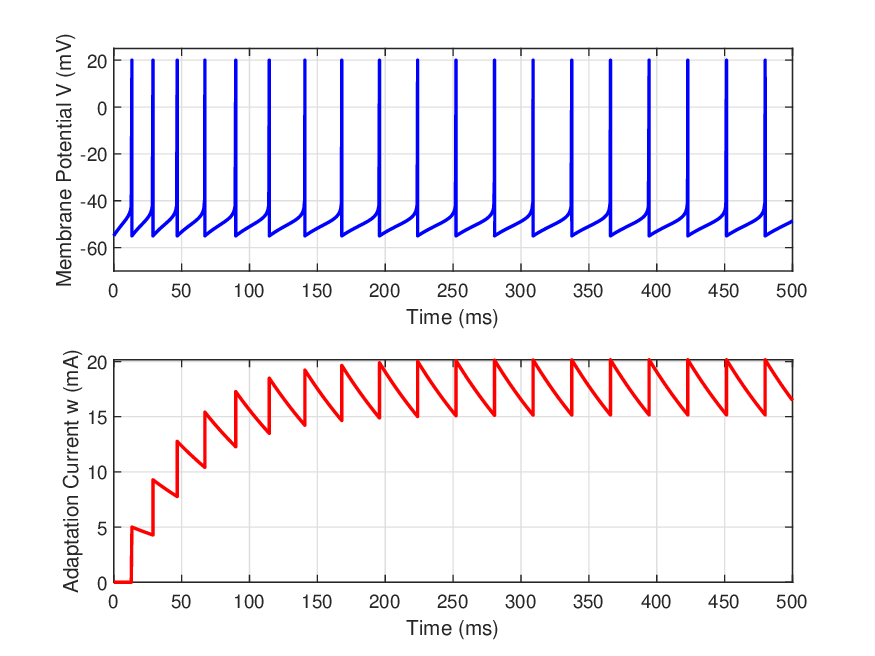}
    \caption{Simulation results of the Adapting pattern in the AdEx model. The upper figure shows the variation of the membrane potential \( V \) over time in the Adapting pattern, while the lower figure shows the variation of the adaptation current \( \omega \) over time in the  Adapting pattern.}
    \label{fig:Adapting pattern}
\end{figure}
\begin{figure}
    \centering
    \includegraphics[width=1.0\linewidth]{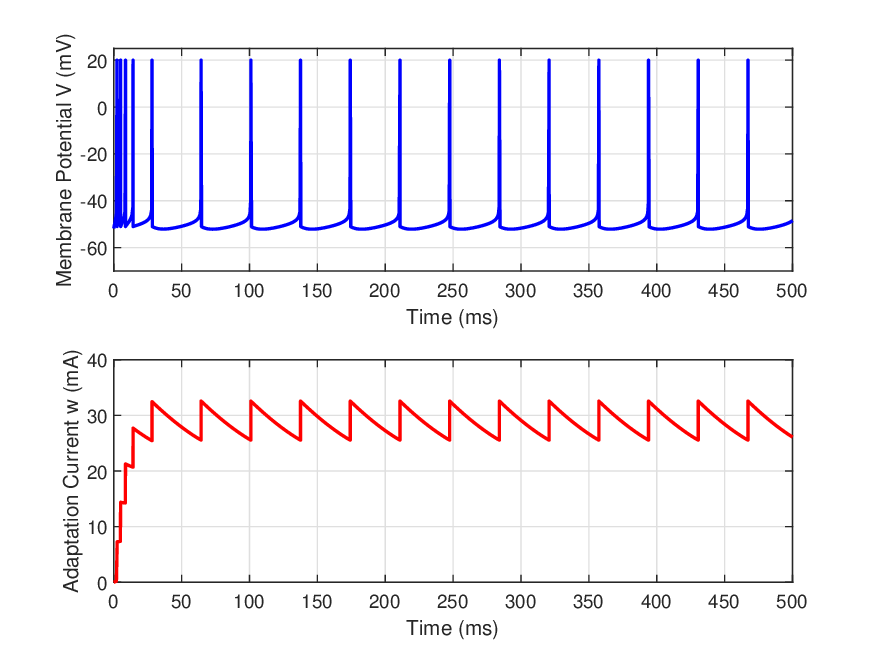}
    \caption{Simulation results of the Initial Busting pattern in the AdEx model. The upper figure shows the variation of the membrane potential \( V \) over time in the Initial Busting pattern, while the lower figure shows the variation of the adaptation current \( \omega \) over time in the Initial Busting pattern.}
    \label{fig:Initial Busting pattern}
\end{figure}
\begin{figure}
    \centering
    \includegraphics[width=1.0\linewidth]{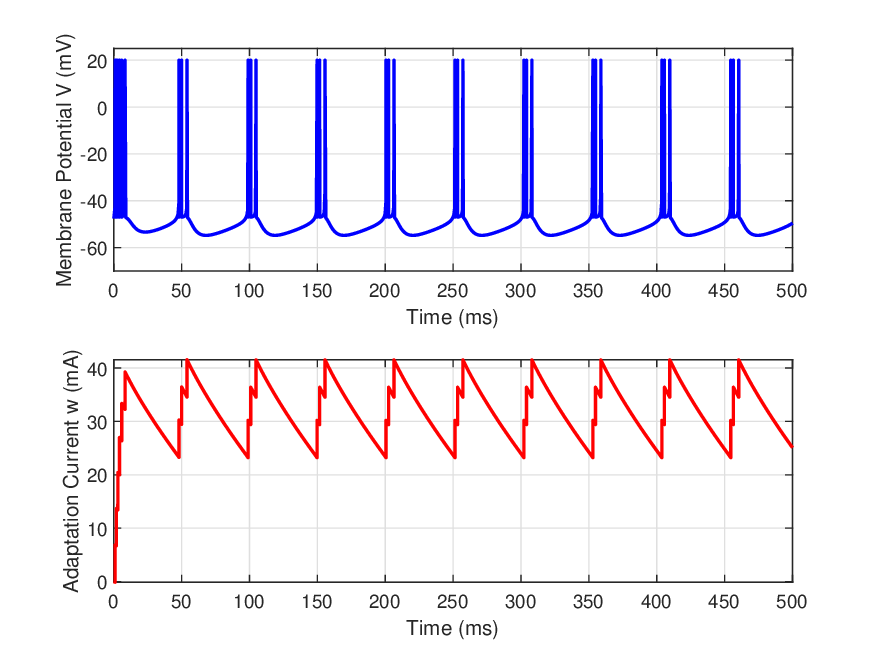}
    \caption{Simulation results of the Busting pattern in the AdEx model. The upper figure shows the variation of the membrane potential \( V \) over time in the Busting pattern, while the lower figure shows the variation of the adaptation current \( \omega \) over time in the Busting pattern.}
    \label{fig:Busting pattern}
\end{figure}
\begin{figure}
    \centering
    \includegraphics[width=1.0\linewidth]{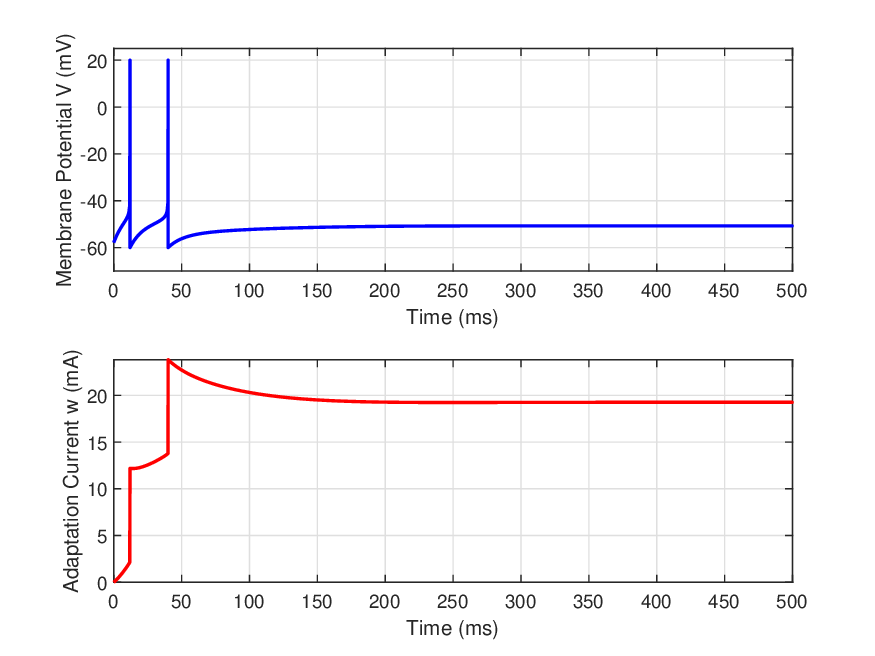}
    \caption{Simulation results of the Transient Spiking pattern in the AdEx model. The upper figure shows the variation of the membrane potential \( V \) over time in the Transient Spiking pattern, while the lower figure shows the variation of the adaptation current \( \omega \) over time in the Transient Spiking pattern.}
    \label{fig:Transient Spiking pattern}
\end{figure}
\begin{figure}
    \centering
    \includegraphics[width=1.0\linewidth]{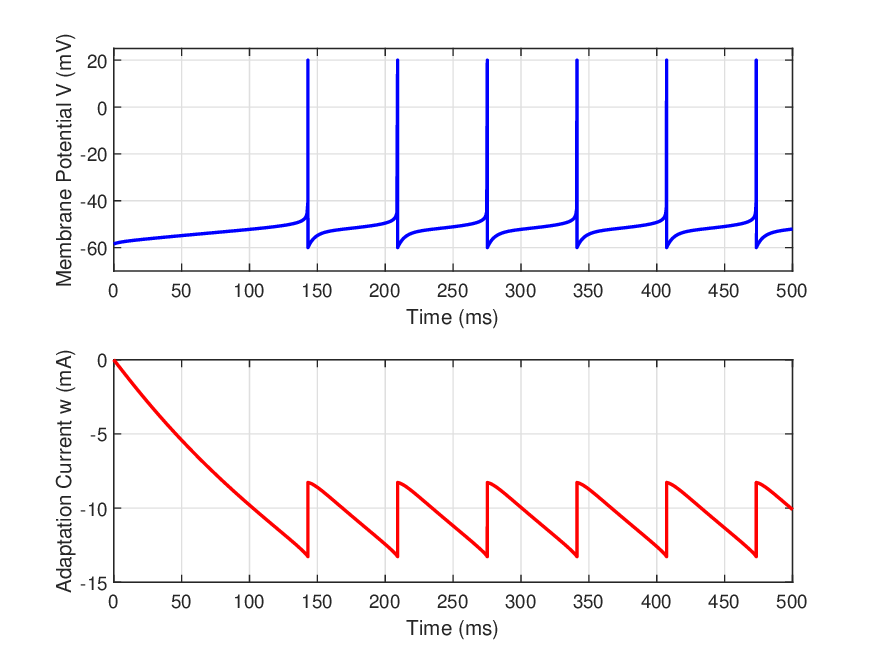}
    \caption{Simulation results of the Delayed Spiking pattern in the AdEx model. The upper figure shows the variation of the membrane potential \( V \) over time in the Delayed Spiking Spiking pattern, while the lower figure shows the variation of the adaptation current \( \omega \) over time in the Delayed Spiking Spiking pattern.}
    \label{fig:Delayed Spiking Spiking pattern}
\end{figure}

Building upon the AdEx framework, we systematically reproduced these firing patterns through numerical simulations. By strategically modulating parameters specified in Table \ref{table1} while maintaining invariant parameters from Table \ref{table2}, we generated six representative firing modalities illustrated in FIGs \ref{fig:Spiking pattern},\ref{fig:Adapting pattern},\ref{fig:Initial Busting pattern},\ref{fig:Busting pattern},\ref{fig:Transient Spiking pattern},and\ref{fig:Delayed Spiking Spiking pattern}. This parametric approach demonstrates the AdEx model's capacity to capture the rich spectrum of neurocomputational dynamics observed in biological systems.

\section{Dynamic Mechanisms of Different Firing Patterns in the Adaptive Exponential Integrate-and-Fire (AdEx) Model with Phase Plane Analysis}

In this section, we discuss the firing mechanism of the six representative firing patterns by phase plane analysis.

\subsection{The Tonic Spiking Pattern}

\begin{figure}
    \centering
    \includegraphics[width=1.0\linewidth]{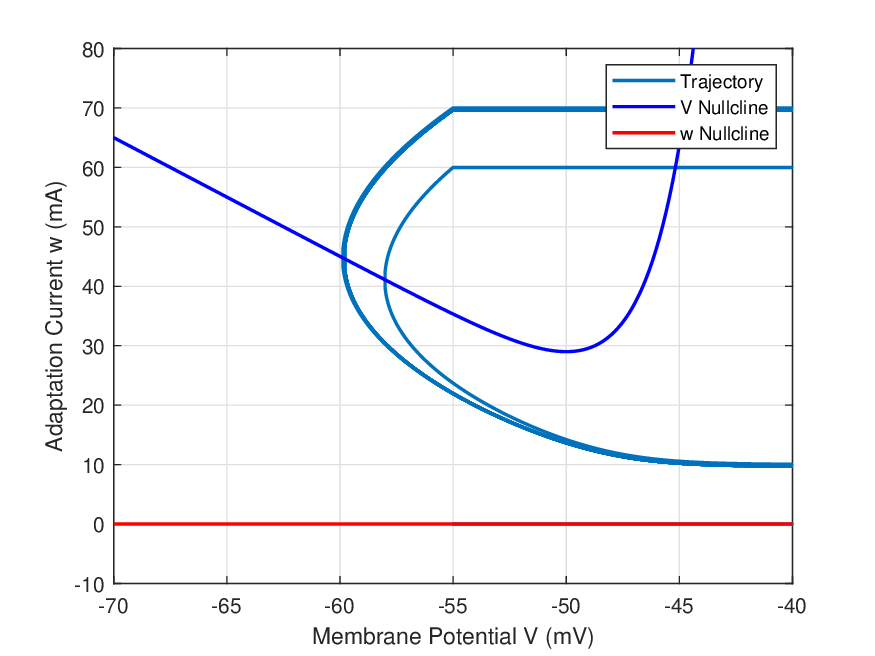}
    \caption{Phase plane analysis diagram of the Tonic Spiking pattern.The phase plane diagram for Tonic Spiking in the AdEx model shows periodic trajectories with initial hyperpolarization and subsequent depolarization due to a large b value (b = 60). This detour across the V-nullcline creates extended inter-spike intervals, preventing rapid successive firing.}
    \label{fig:Tonic Spiking phase}
\end{figure}

The phase plane diagram for Tonic Spiking pattern in the AdEx model (FIG.\ref{fig:Tonic Spiking phase}) reveals periodic trajectory motion that generates the Tonic Spiking pattern. After each reset of membrane potential V and adaptation current w, the state point follows a circuitous path before triggering an action potential, manifesting as an initial decrease in membrane potential followed by a subsequent increase.

This behavior results from the large b parameter value (b = 60 in this paper(Table\ref{table1})), which positions the post-reset state below the V-nullcline. Critical dynamics emerge from the sign reversal of $dV/dt$ across this boundary: above the V-nullcline, $dV/dt < 0$ produces hyperpolarization, while below it, $dV/dt > 0$ enables depolarization. Consequently, the trajectory must first traverse the hyperpolarizing region (where V decreases) until crossing into the depolarizing zone below the nullcline.

This obligatory detour creates extended inter-spike intervals, fundamentally preventing rapid successive firing due to the dominant -w term in the membrane potential equation immediately following reset, where substantial adaptation accumulates through large b increments.

\subsection{The Adapting Pattern}

\begin{figure}
    \centering
    \includegraphics[width=1.0\linewidth]{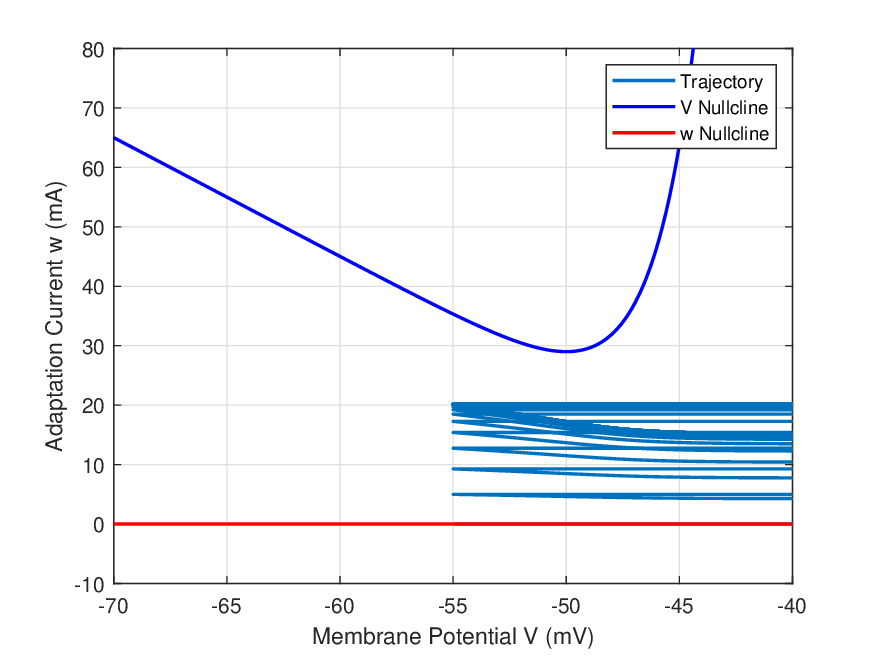}
    \caption{Phase plane analysis diagram of the Adapting pattern.Adapting pattern has increasing inter-spike intervals due to rising \( w \) reset points. When reset increments match previous \( w \) reductions, intervals stabilize. Unlike tonic spiking, adaptive firing trajectories avoid the \( V \)-nullcline, directly depolarizing from \( V_{\text{reset}} \).}
    \label{fig:Adapting phase}
\end{figure}

The phase plane diagram for adaptive firing in the AdEx model is illustrated in FIG. \ref{fig:Adapting phase}. After each spike, the reset point of adaptation variable \( w \) is higher than the previous one. The time required for the trajectory point to travel from its starting position to spike generation gradually increases, manifesting as progressively lengthening inter-spike intervals. When the vector field causes membrane potential to descend further along the \( w \)-axis, if the increment \( b \) in the current reset equals the reduction in \( w \) during the previous trajectory—meaning the current reset point coincides with the previous reset point—\( w \) ceases to increase further. At this stage, the neuron's inter-spike interval stabilizes to a fixed value. This constitutes the dynamic mechanism underlying adaptive behavior in the AdEx model.

The primary distinction between adaptive firing and tonic spiking lies in the trajectory never crossing the \( V \)-nullcline. This occurs because, in adaptive firing, parameter \( b \) takes a relatively small value (e.g., \( b = 5 \)), resulting in minimal increments of \( w \) during resets. Consequently, neurons exhibiting adaptive behavior bypass the subthreshold process of initial membrane potential \( V \) decrease followed by increase. Instead, they depolarize directly from the reset point \( V_{\text{reset}} \) until reaching action potential threshold.

\subsection{The Initial Bursting Pattern}

\begin{figure}
    \centering
    \includegraphics[width=1.0\linewidth]{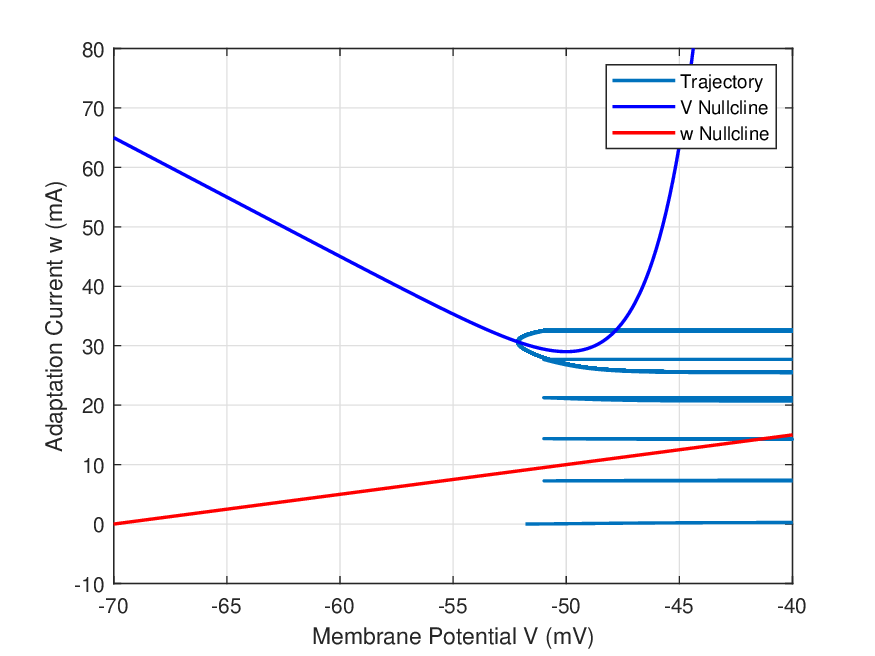}
    \caption{Phase plane analysis diagram of the Initial Bursting pattern. In Initial Bursting in the AdEx model, neurons first fire high-frequency bursts due to large adaptation increments, then transition to regular firing as adaptation increases and membrane potential decreases.}
    \label{fig:Initial Bursting phase}
\end{figure}

The phase plane diagram for initial bursting in the AdEx model is illustrated in FIG. \ref{fig:Initial Bursting phase}. In initial bursting mode, the motion of trajectory points can be divided into two primary phases. First is the initial bursting phase. After a neuron fires an action potential, the membrane potential resets to the reset potential while the adaptation variable increases by a large fixed increment. Due to this substantial increment, the post-reset point typically falls below the membrane potential's nullcline. Below this nullcline, the membrane potential rapidly rises, causing the trajectory point to move in the upper-right direction. This results in a series of high-frequency action potentials, forming a burst.  

Next is the adaptation phase. As the adaptation variable increases, the trajectory point gradually approaches and crosses the membrane potential's nullcline. At this stage, the membrane potential begins to decrease, and the trajectory point moves in the lower-left direction. During this phase, the decline in membrane potential combined with the slow decay of the adaptation variable reduces firing frequency, transitioning the neuron into regular or adaptive firing mode.  

Parameter \( b \) plays a critical role in initial bursting. Larger \( b \) values cause post-reset trajectory points to land below the membrane potential's nullcline, triggering initial bursting. Subsequently, the accumulation of the adaptation variable forces the trajectory into a circuitous path, ultimately entering the adaptation phase. Conversely, smaller \( b \) values may place post-reset points above the nullcline, leading to rapid membrane depolarization and immediate re-firing. In such cases, the neuron enters regular firing mode without exhibiting distinct bursting behavior.

\subsection{The Bursting Pattern}

\begin{figure}
    \centering
    \includegraphics[width=1.0\linewidth]{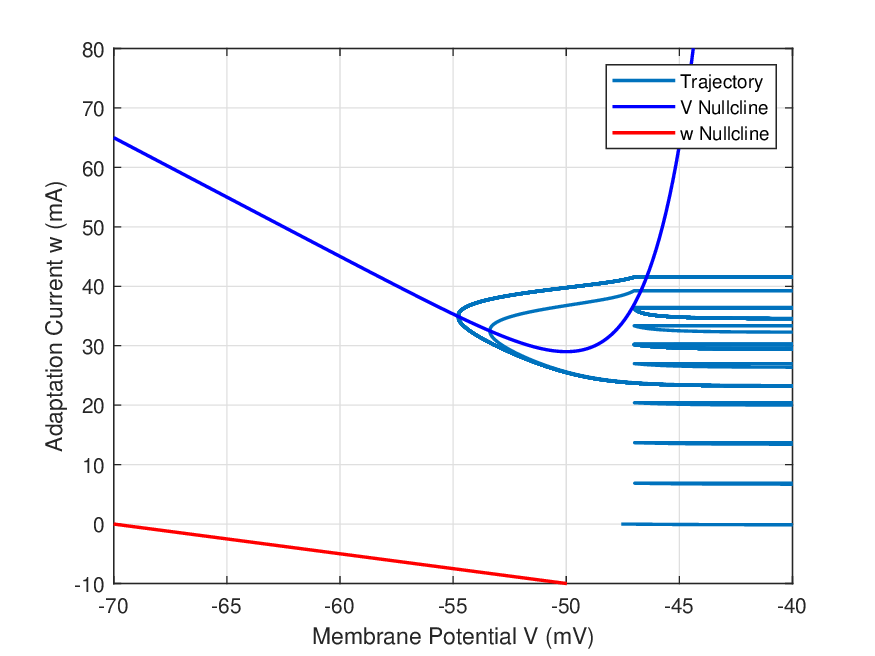}
    \caption{Phase plane analysis diagram of the Bursting pattern. In Bursting mode of the AdEx model, neurons cycle through high-frequency bursts, quiet interburst intervals, and preparation phases. Larger \( b \) values promote bursting, while larger \( a \) values prolong interburst intervals.}
    \label{fig:Bursting phase}
\end{figure}

The phase plane diagram for bursting in the AdEx model is illustrated in FIG.  \ref{fig:Bursting phase}. In bursting mode, the motion of trajectory points can be divided into several stages. First is the bursting phase. After a neuron fires an action potential, the membrane potential resets to the reset potential while the adaptation variable increases by a large fixed increment. Due to this substantial increment, the post-reset point typically falls below the membrane potential's nullcline. Below this nullcline, the membrane potential rapidly rises, causing the trajectory point to move in the upper-right direction. This results in a series of high-frequency action potentials, forming a burst.

Next is the interburst interval. As the adaptation variable increases, the trajectory point gradually approaches and crosses the membrane potential's nullcline. At this stage, the membrane potential begins to decrease, and the trajectory point moves in the lower-left direction. During the interburst interval, the decline in membrane potential combined with the slow decay of the adaptation variable reduces firing frequency, causing the neuron to enter a relatively quiescent period.

This is followed by the preparation phase for the next burst. During the interburst interval, the adaptation variable gradually decays, causing the trajectory point to slowly move in the upper-left direction. When the adaptation variable decays sufficiently, the trajectory point again approaches the membrane potential's nullcline, initiating a new cycle of rapid membrane potential rise in preparation for generating the next burst.

Parameter \( b \) plays a critical role in bursting. Larger \( b \) values cause post-reset trajectory points to land below the membrane potential's nullcline, thereby triggering bursting. The magnitude of \( b \) determines the number of action potentials within each burst and the burst duration. Another crucial parameter is \( a \), which governs voltage sensitivity of the adaptation current. Larger \( a \) values enhance adaptation effects, prolonging interburst intervals and increasing the duration between bursts.

\subsection{The Transient Spiking Pattern}

\begin{figure}
    \centering
    \includegraphics[width=1.0\linewidth]{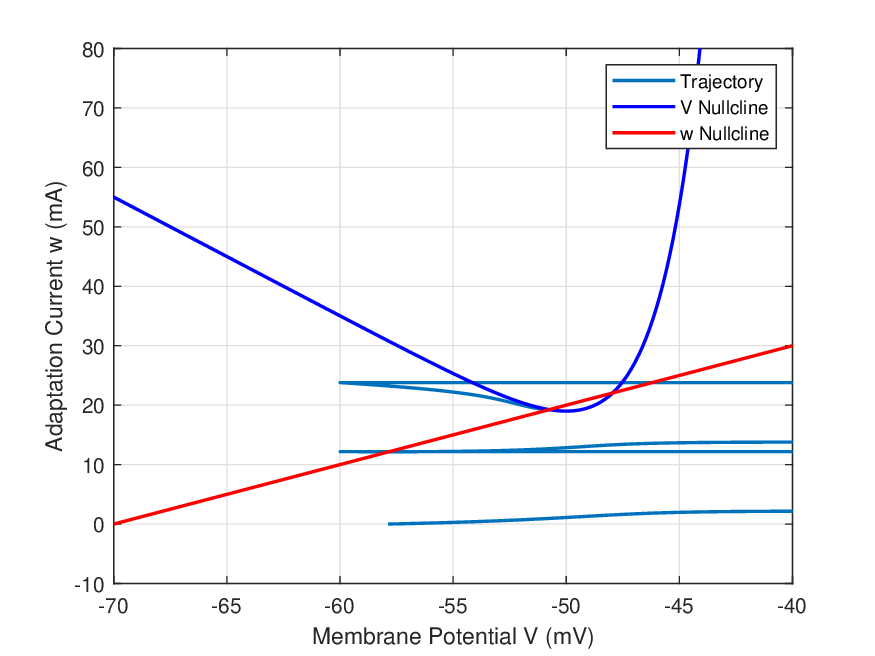}
    \caption{Phase plane analysis diagram of the Transient Spiking pattern.In transient spiking in the AdEx model, initial small adaptation variable \( w \) causes the neuron to fire repeatedly until the trajectory enters the stable focus, after which firing ceases.}
    \label{fig:Transient Spiking phase}
\end{figure}

The phase plane diagram for transient spiking in the AdEx model is illustrated in FIG. \ref{fig:Transient Spiking phase}. During transient spiking, the dynamic behavior of membrane potential can be understood through phase plane analysis. Initially, due to the small value of adaptation variable \( w \), the neuron's starting state is distant from the stable focus. At this stage, membrane potential continuously rises and triggers action potentials. This pattern repeats until the post-reset state point enters the basin of attraction of the stable focus. Once the trajectory point is drawn toward the stable focus, both membrane potential and adaptation variable \( w \) will asymptotically approach stability, causing the neuron to cease firing action potentials.

\subsection{The Delayed Spiking Parttern}

\begin{figure}
    \centering
    \includegraphics[width=1.0\linewidth]{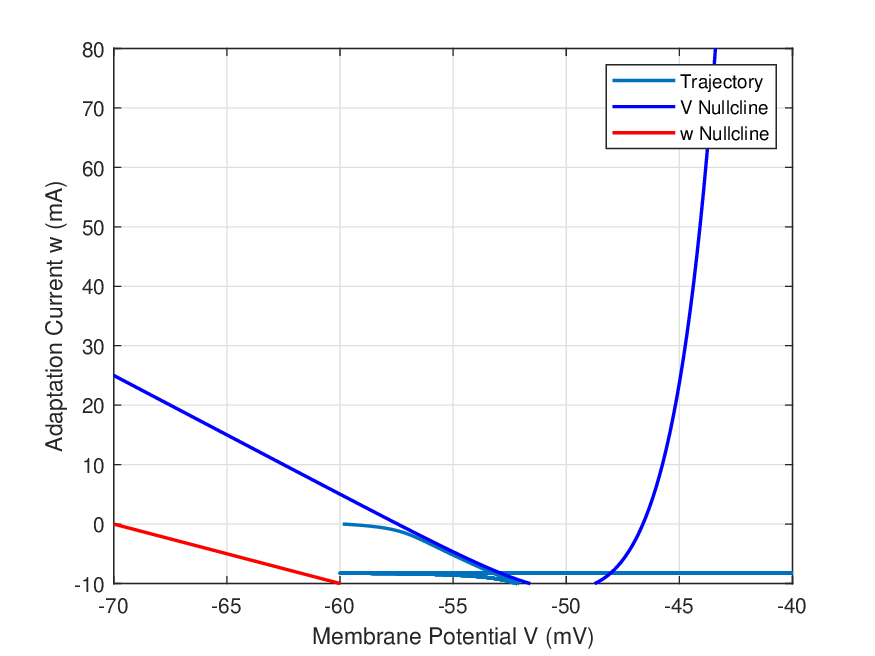}
    \caption{Phase plane analysis diagram of the Delayed Spiking pattern.In delayed spiking in the AdEx model, the trajectory moves extremely slowly near the membrane potential's nullcline, causing the neuron to take a long time to accumulate enough potential to fire its first action potential.}
    \label{fig:Delayed Spiking phase}
\end{figure}

The phase plane diagram for delayed spiking in the AdEx model is illustrated in FIG. \ref{fig:Delayed Spiking phase}, showing extremely slow motion of the trajectory during the initial phase. This occurs because when the trajectory approaches the membrane potential's nullcline, its horizontal velocity component becomes very small. Points on the nullcline satisfy \( dV/dt = 0 \), meaning membrane potential exhibits no horizontal movement—only vertical changes. Although trajectory points near the nullcline still display horizontal motion, their minimal velocity component results in extremely slow overall movement.

This sluggish progression causes the neuron to require extended time to accumulate sufficient membrane potential before firing its first action potential. Only when membrane potential ultimately reaches adequately high levels can the trajectory rapidly cross the nullcline and trigger an action potential. Therefore, delayed firing formation relies fundamentally on this slow movement near the nullcline, a characteristic feature that dictates the neuron's prolonged accumulation period before spiking.

\section{Conclusion}

In this paper,we employ phase plane analysis to conduct a depth investigation of diverse firing patterns in the AdEx model, revealing their underlying dynamic mechanisms. Through systematic analysis of Tonic Spiking, Adapting , Initial Bursting, Bursting, Transient Spiking, and Delayed Spiking firing patterns, we comprehensively examine the characteristic trajectory features and formation mechanisms of each pattern within the phase plane. These findings not only enrich our understanding of neuronal firing patterns but also provide novel theoretical support for neuroscience research.

The principal insight gained in this paper is that the AdEx model can simulate multiple neuronal firing patterns through parametric modulation, with each pattern exhibiting distinct dynamic mechanisms. These discoveries hold significant implications for understanding neuronal behaviors under varying physiological conditions, with potential applications in neural network modeling, neurological disorder research, and brain-computer interfaces. Future researches could explore firing patterns across broader parametric combinations and conduct more rigorous model validation and optimization through integration with experimental data.

\begin{acknowledgments}
    We extend special gratitude to Professor Xiong-Fei Wang , whose meticulous guidance and invaluable suggestions during this semester's Scientific Paper Writing course played a pivotal role in shaping the structural design of this paper. Furthermore, We express appreciation to Associate Professor Lian-Chun Yu and Professor Dong-Sheng Lei , instructors of Computational Physics II last semester, whose instruction laid the foundational basis for this research. Finally, We acknowledge the authors of Neural Computation Modeling Practical: Based on BrainPy\cite{brainpy}, as this book ignited our interest in computational neuroscience while the work of this paper constitutes a learning outcome from studying its contents.
\end{acknowledgments}

\bibliographystyle{plain} 
\bibliography{apssamp} 

\end{document}